\documentclass[11pt,oneside,letterpaper]{article}
\usepackage{amssymb}
\usepackage{amsmath}
\usepackage[dvips]{graphicx}
\usepackage{setspace}
\usepackage{amsfonts}
\usepackage{fancyhdr}
\usepackage{xcolor}
\usepackage{graphicx}
\usepackage{rotating}
\usepackage{comment}
\usepackage{subfigure}
\usepackage{color}
\usepackage{cite}
\usepackage{braket}
\usepackage{caption}
\usepackage[utf8]{inputenc}
\usepackage{bbold}
\usepackage[export]{adjustbox}
\usepackage[bookmarks=false]{hyperref}

\definecolor{darkgreen}{rgb}{0,0.5,0}
\definecolor{darkblue}{rgb}{0,0,0.6}
\definecolor{purple}{rgb}{0.4,.2,0.7}

\newcommand{\be}{\begin{equation}}
\newcommand{\ee}{\end{equation}}

\usepackage{pdfsync}
\makeatletter
\newcommand*{\defeq}{\mathrel{\rlap{%
                     \raisebox{0.3ex}{$\m@th\cdot$}}%
                     \raisebox{-0.3ex}{$\m@th\cdot$}}%
                     =} 
\makeatother

\def\be{\begin{eqnarray}}
\def\ee{\end{eqnarray}}

\newcommand{\bea}{\begin{eqnarray}}
\newcommand{\eea}{\end{eqnarray}}
\def\ben{\begin{equation}}
\def\een{\end{equation}}

   \let\k=\kappa
     \let\r=v

\def\be{\begin{equation}}
\def\ee{\end{equation}}
\def\ba{\begin{array}}
\def\ea{\end{array}}

\def\ba#1\ea{\begin{align}#1\end{align}}
\def\bs#1\es{\begin{split}#1\end{split}}

\newcommand{\tp}{\tilde \Psi}

\allowdisplaybreaks  % allow page breaks in displayed eqs

\numberwithin{equation}{section}
\def \be {\begin{equation}}
\def \ee {\end{equation}}
	
\def \JM#1 {{\color{blue}  JM: #1 }}
\def \AAl#1 {{\color{red}  AA: #1 }}

\begin{document}
\onehalfspacing

\begin{center}

~
\vskip5mm

{\LARGE  {
Glimmers of a post-geometric perspective
\\
\ \\
}}

Federico Piazza

\vskip5mm
{\it Aix Marseille Univ, Universit\'{e} de Toulon, CNRS, CPT, Marseille, France
} 

\vskip5mm

{\tt piazza@cpt.univ-mrs.fr }

\end{center}

\vspace{4mm}

\begin{abstract}
\noindent
Quantum gravitational effects can become important at low energy if the wavefunction of the metric field fails to be peaked around a classical configuration. 
We try to understand such deviations from classicality within canonical quantum gravity by introducing a ``fluid of observers" in the low energy theory and defining a distance operator ``at equal time" among them. 
We find that, even in the presence of relevant fluctuations in the metric field, a ``locally flat" limit is recovered in the neighbourhood of each observer. Deviations from classicality have no particular consequence, locally. However, at larger separations the expectation value of the distance operator behaves differently than a standard Riemannian distance. In particular, it is non-additive and thus cannot be obtained by the integral of a differential line element. 
This emerging ``beyond Riemannian" geometry is a metric space similar to embedded Riemannian manifolds equipped with \emph{chord} distances that ``cut through" the ambient space. We study deviations from flat space by looking at triangles in the limit where their sizes go to zero. Beyond-Riemannian deviations with respect to flat space are of the same order as standard Riemannian ones, but qualitatively different. Possible connections with holography and with the black hole information paradox are briefly discussed. 

 \end{abstract}
%\vspace{.2in}
%\vspace{.3in}

\pagebreak
\pagestyle{plain}

\setcounter{tocdepth}{2}
{}
\vfill

\ \vspace{-2cm}
\renewcommand{\baselinestretch}{1}\small
\tableofcontents
\renewcommand{\baselinestretch}{1.15}\normalsize

\section{Introduction}

Spacetime is portrayed in general relativity as a (pseudo-) Riemannian manifold. 
What are the limits of such a classical description? Quantum gravitational effects must become important at  least when curvature invariants are of order Planck as, for example, in the regions approaching a singularity.  
However, aspects of quantum gravity could be at play already in the low energy theory and at macroscopic scales,  if the wavefunction of the metric field fails to be peaked around any classical configuration.

The classicality of the wavefunction is a known issue in quantum cosmology (see e.g.~\cite{Halliwell:1990uy}), where one usually assumes a strong correlation among the canonical variables so that a classical metric can emerge as a coherent state in the appropriate limit. 
However, such a correlation is not guaranteed to exist nor to be preserved~\cite{Dvali:2017eba,Berezhiani:2020pbv,Dvali:2020etd} during time evolution. Typical examples of highly quantum states are those of a system undergoing tunneling. 
The use of instanton methods for bubble nucleation~\cite{Coleman:1980aw} tends to conceal that between the two classical configurations that the instanton interpolates spacetime is not defined in any classical sense. In more general situations one could imagine a tunneling that does not even reach a classical configuration eventually. Depending on the characteristic time scales, rather than a pathology this could simply represent a peculiar quantum behavior that cannot be captured by classical evolution. 
We speculate that such a non-classical behavior of the gravitational field is a general feature of certain time dependent situations (see Sec.~\ref{sec_constr} below) whose complete and accurate description should thus include deviations from the classical geometrical picture of general relativity. 

We attempt to better understand these states in canonical quantum gravity. Together with the three-dimensional metric, the positions of a set of observers which serve as a reference frame are the main canonical variables. The corresponding eigenvectors represent classical three-geometries with fields and observers perfectly defined. On this basis one can calculate distances between observers with the standard geodesic prescription, which allows to formally define a distance operator in the low energy theory at a non-perturbative level. 
Although dynamical reference frames have already been discussed in quantum gravity (e.g.~\cite{Rovelli:1990ph,Brown:1994py}), the idea of a distance operator is new, to the best of our knowledge. The expectation value of the distance operator displays a locally flat limit like standard Riemannian distances but anomalous ``\emph{beyond-Riemannian}" properties at large separations.

In the rest of this section we provide motivations for this research. 
We dedicate Sec.~\ref{sec_2new} to a more in depth summary and plan of the rest of the paper.

\subsection{Why are you doing this?}

Quantum gravity is best understood and under control in asymptotically anti-de Sitter space, where one can benefit of the insights of the AdS/CFT correspondence~\cite{Maldacena:1997re}. Classical spacetime emerges in this setup as the saddle point approximation for quantities defined on the boundary. Notably, fine-grained Von Neumann entropies can be calculated with Euclidean path integrals already at the level of the low-energy effective theory~\cite{Lewkowycz:2013nqa,Faulkner:2013ana}. By using these tools a number of breakthroughs on the black hole information paradox have been obtained and the Page curve of Hawking radiation correctly reproduced (see e.g.~\cite{Jafferis:2017tiu,Almheiri:2020cfm,Raju:2020smc,Almheiri:2019psf,Penington:2019npb} and references therein). 
Different Euclidean spacetimes~\cite{Hawking:1982dh} and different topologies in the bulk~\cite{Penington:2019kki,Almheiri:2019qdq} contribute to these calculations.  These sparkling Euclidean constructions, however,  leave the Lorentzian side of the picture relatively  in the dark. How to reconstruct the actual Lorentzian spacetime, say, as experienced by an observer falling into the black hole\footnote{This operational point of view has been undertaken e.g. in~\cite{Jafferis:2020ora}}? 
More generally, Euclidean saddles seem far more telling and efficient computational tools than Lorentzian ones.  How does the information about some measurement happening inside the black hole eventually evade to the outside without violating causality? We argue that some of these struggles, along with other puzzles recently discussed in~\cite{Geng:2021hlu}, are rooted in insisting to stage black hole evaporation in a classical Lorentzian spacetime.

Another inspiration for the present work comes from a promising approach to bulk reconstruction, that of integral geometry~\cite{Czech:2015qta}. The idea is to recover a geometry from the lengths of the geodesics living therein. In asymptotically AdS spaces of arbitrary dimension one should consider the set of minimal codimension two surfaces anchoring to the AdS boundary. The authors of~\cite{Czech:2015qta} concentrate on AdS$_3$, where such surfaces are in fact geodesics. One can assign a length to these curves by calculating the Von Neuman entropy of the subtended spatial regions of the CFT and applying the Ryu Takayanagi formula~\cite{Ryu:2006bv,Hubeny:2007xt}. Without all subtleties related to quantum gravity, and for arbitrarily ``highly quantum" states, the entropy of a boundary region is always a well defined QFT quantity. The corresponding bulk lengths can thus be taken as the fundamental building blocks to reconstruct the metric. 
 But what if such lengths are not supported by any Riemannian geometry? 
Clearly, a metric tensor can be used to calculate all geodesic distances between points. But randomly chosen distances between points cannot always be obtained as the integral of a differential line element (see the rest of the paper for examples). 
In this case one would have to deal with a ``beyond-Riemannian" bulk geometry. 

One last motivation for venturing beyond classical geometry is provided by cosmology. 
After years of successful confirmations, data of increasing precision have stopped converging so enthusiastically towards a single $\Lambda$CDM model. The disagreement between the direct and indirect measurements of the Hubble constant~\cite{Riess:2020sih,DiValentino:2021izs} looks difficult to fix  with just minor adjustments and might be calling for a deeper reassessment of our basic geometrical description of the universe. Among other (milder) inconsistencies~\cite{Perivolaropoulos:2021jda}, one is particularly smelling  of ``geometrical anomaly", namely the fact that a positive spatial curvature seems preferred by some observables at around $3\sigma$ and  excluded by most others~\cite{Aghanim:2018eyx,DiValentino:2020hov}.
But this is not the end of the story, if the reader can tolerate any more philosophy at this point.  
Accelerating expansion must be invoked~\emph{twice} within the standard geometric framework, at disparate epochs and disparate energy scales. The whole evolution of the universe is sandwiched in between two accelerations, inflation and dark energy. Such a dissociated picture is by now very familiar and accepted among professional cosmologists. 
Nonetheless, it remains a curious circumstance, that some alternative geometric framework could perhaps unify away. 

\subsection{Constraints and range of applicability}\label{sec_constr}

In this paper we suggest that a beyond-Riemannian geometry could indeed emerge in quantum gravity by studying the expectation value of a \emph{spatial} distance operator, defined on suitable ``equal time surfaces". Our approach is kinematical because we consider from the onset states with relevant metric fluctuations.
The \emph{dynamical} question to address next would be: in which low energy regimes (if any) could the theory display such relevant deviations from classicality? Here follow some qualitative/phenomenological considerations.

Static or quasi static backgrounds provide the most severe experimental constraints to  any beyond-Riemannian ambition.
The satellite-based global positioning system (GPS) on earth together with the many earth-moon and solar system tests include direct measurements of lengths and time intervals which show exquisite agreement with classical general relativity. 
If macroscopic quantum gravity effects are at play in this world, they better be related with time evolution and die off in quasi-static situations. Successful gravitational wave experiments---and our full understanding of the underlying dynamics---extend this no-go zone to those time-dependent environments  where on-shell gravitons---or Weyl curvature invariants---account for most of the dynamics. Time-varying  Ricci  curvatures are far less tested. One example is the interior of a collapsing star and the spacetime around an evaporating black hole on which the matter content of Hawking radiation back-reacts.  The main other  notable example is cosmology. The expanding universe entails important time variations of the Ricci curvature over Hubble scales. As argued before, this is a realm where macroscopic quantum gravity effects, while not completely excluded, may actually be longed for.

\section{Main ideas and plan of the paper}\label{sec_2new}

One notorious difficulty in quantum gravity is that of defining gauge invariant quantities. 
In practice, as we aim to define a distance operator, we should first understand  \emph{between what and what} we want to measure distances. The problem can be posed on firmer grounds by introducing a set of dynamical objects that can serve as a reference frame. A very natural type of reference is a certain number of observers/test-particles following their worldlines. In the next sections of the paper we will opt for an improved and mathematically smoother version of this, namely a continuum ``fluid of observers". Here however we convey the main ideas with the more intuitive test-particles picture. 

\subsection{Observers}

The test particles that we include in our system are actually real observers, recognizable and distinguishable from each other. Instead of calling them e.g. Charlie, Mick, Keith etc. we will name them by capital latin letters for notational convenience. They look like points, or worldlines, just because we deal with them at lowest order in derivatives in a low energy approximation~\cite{Goldberger:2004jt},
\begin{equation} \label{2-intro}
S  =  \frac{1}{16 \pi G}\int d^4 x \sqrt{-g} R  \ - \  m_{A} \!\int d\tau_{A} \ - \ m_{B} \!\int d\tau_{B} \  + \ \dots\, ,
\end{equation}
where ellipses stand for other observers, finite size effects on the worldlines and matter fields not further specified. The canonical ADM variables of this system are the spatial metric $h_{ij}(x^k)$ and the spatial positions\footnote{They are contained inside the worldline actions because
$$
d \tau_{A} = dt \sqrt{g_{\mu \nu} \frac{dx^\mu_{A}}{dt} \frac{d x^\nu_{A}}{dt}}\, , \quad {\rm etc.}
$$}
 of the observers $x^i_{A}$, $x^i_{B}$, etc. on some given time foliation, together with their conjugate momenta. 

We aim for a low energy---and still non perturbative---description of this system. 
We thus consider states that are functionals of the metric field and functions of the positions, 
\begin{equation}
\Psi = \Psi\left[h_{ij}(x^k), \ x^i_{A}, \ x^i_{B}, \ \dots\right]\, .
\end{equation}
Now we know \emph{between what and what} we want to measure the distance---between $A$ and $B$! But the problem of gauge invariance is not  completely solved yet. We still have to deal with time evolution.  
Gauge invariant distances imply the full four-dimensional extremisation of the length of a curve between two points/events. 
Even when $\Psi$ is peaked on a classical configuration, it displays the system on a time slice that is chosen arbitrarily. We cannot use such a spatial configuration to calculate any meaningful gauge invariant distance. 

\subsection{Time}

There is a trick for fixing the time slice in a gauge invariant way.  Let us introduce a dynamical scalar field $T(x^k)$ (see Sec.~\ref{sec_timefield} for more details) and  extract from the full state $\Psi$ a \emph{conditional probability amplitude} $\tp$ for all other fields. The condition to satisfy here is that $T(x^k)$ be equal to some constant value $T_0$,
\begin{equation} \label{wave-func}
\tilde \Psi\left[h_{ij} (x^k), \, \ x^i_{A}, \ \dots ;\ T_0\right] \ \equiv\ {\cal N}  \Psi\left[h_{ij} (x^k), \, \ x^i_{A}, \ \dots  \, , \  T(x^k) =T_0\right] , \
\end{equation}
 The ``state" $\tp$ is a fair description of the system ``at a given time". Thanks to the scalar field $T$  the surface at constant time is gauge invariantly defined and we can measure distances on it. Other subtleties related with the momentum and Hamiltonian constraints are dealt with in Sec.~\ref{sec_timefield} and will not be discussed further in this summary. 

\subsection{Classical basis}
The next problem is how to define a distance in the presence of relevant departures from classicality.  To this aim we can expand  $\tp$ on a ``\emph{classical basis}" of states $|h_{ij}(x^\k), x^i_{A},\dots\rangle$ i.e. on a basis of common eigenvectors of the commuting operators $\hat h_{ij}$, $\hat x^i_{A}$, etc.\footnote{Of course, the wavefuntional~\eqref{wave-func} \emph{is} already the scalar product of the state $|\tp\rangle$ with the  generic element of this basis, i.e. 
$$ 
\tp[h_{ij}(x^k), \ x^i_{A},  \ \dots; T_0]\ \equiv \ \langle h_{ij}(x^\k), x^i_{A},\dots | \tp;T_0\rangle\, .
$$}
On each element of such a basis the distance between $A$ and $B$ is defined with infinite accuracy.  One has to find the geodesic connecting $x_A$ and $x_B$ in the corresponding classical geometry and calculate its length.  Let us call $ d({A}, {B})_{\{h, x_{A}, x_B \}}$ the result of this computation. We can define the distance operator between $A$ and $B$ by its action on this classical basis,  
\begin{equation}\label{classbasis}
\hat d({A}, {B}) \ |h_{ij}(x^\k), x^i_{A},\dots\rangle =  d({A}, {B})_{\{h, x_{A}, x_{B}\}} \ |h_{ij}(x^\k), x^i_{A},\dots\rangle\, .
\end{equation}
In other words, we define the distance operator by declaring that the elements of the classical basis are its eigenvectors, of eigenvalues $d({A}, {B})_{\{h, x_{A}, x_{B}\}}$. 

\subsection{Average and beyond-Riemannian distances}
By linear superposition we can compute the expectation value of the distance operator between $A$ and $B$ on any state $\tp$,
\begin{equation} \label{avedist}
\langle \hat d({A}, {B}) \rangle = 
\int {\cal D}  h\  d^3  x_{A} \  d^3 x_{B} \ d^3 x_C\dots \ \left |\langle h_{ij}(x^\k), x^i_{A},\dots | \tp;T_0\rangle\ \right |^2  d({A}, {B})_{\{h, x_{A}, x_{B}\}}\, .
\end{equation}

For reasons of analyticity, however, the quantity that should replace the classical notion of distance is probably a different expression, obtained by taking the square root of the expectation value of the \emph{square} of the distance operator. 
\begin{equation}
\overline{d(A,B)} \ \equiv \ \sqrt{\left\langle \hat d(A,B)^2\right\rangle}, 
\end{equation}
where, clearly, 
\begin{equation} \label{avedist2}
\left\langle \hat d({A}, {B})^2 \right\rangle = 
\int {\cal D}  h\  d^3  x_{A} \  d^3 x_{B} \ d^3 x_C\dots \ \left |\langle h_{ij}(x^\k), x^i_{A},\dots | \tp;T_0\rangle\ \right |^2  \left(d({A}, {B})_{\{h, x_{A}, x_{B}\}}\right)^2\, .
\end{equation}
As we need a short name for $\overline{d(A,B)}$ we will call it \emph{beyond-Riemannian distance}, or just \emph{BR-distance}, not to be confused with the average of the distance operator defined in~\eqref{avedist}. 
One main message of this paper is that BR-distances are not additive and thus cannot be expressed as integrals of some ``average differential line element"---although they can in the limit of small separation.  It is quite straightforward to see that averaging over different geometries can produce this effect, and the two examples discussed at the end of Sec.~\ref{sec_avdis} might satisfy the immediate curiosity of the reader. 

We argue in Sec.~\ref{2.1}, on the other hand, that the BR-distance respects a \emph{locally flat} limit in the vicinity of each observer and bears similarity with the chord distance of embedded manifolds. In Sec.~\ref{sec_BR} we attempt to characterize intrinsically the behavior of such distances by considering triangles of small sizes  in the standard Riemannian case first and then in the beyond-Riemannian one.

Finally, in App.~\ref{appen} we will make a last use of the discrete set of observers introduced in this section by setting them in the bulk of an asymptotically AdS space.

\section{Setup}\label{sec_setup}

First we should arm ourselves with a set of fields that can serve as a reference frame. We are going to be dealing with a theory of the type 
\begin{equation} \label{2}
S  \ = \  \frac{1}{16 \pi G}\int d^4 x \sqrt{-g} R \  + \ S_{Ref}[X^I, T] + \ S_m[\phi] \ + \ \dots\, .
\end{equation}
Now the labels of the different observers ($A$, $B$ etc. in~\eqref{2-intro}) are promoted to scalar fields  $X^I$,  $I = 1,2,3$. Each triplet of values for the fields $X^I$ identifies a given observer. So the observers form a continuum. The scalar $T$ is a ``time field" and $\phi$ collectively denotes all other matter fields. In interesting situations one should think of the reference fields as dynamically subdominant components. 

Clearly, the above action is only valid at low energy. In particular, the four reference fields can describe the effective theory of a solid, a fluid or a ``supersolid"~\cite{Son:2005ak}, depending on the internal symmetries and on the symmetry breaking pattern~\cite{Nicolis:2015sra}. The most natural generalization of the discrete model in~\eqref{2-intro} is a non-relativistic fluid of observers\footnote{In this case, at lowest order in derivatives, 
$$S_{Ref}[X^I, T] = \int d^4 x \sqrt{-g} \sqrt{\det B^{IJ}} + S_T[T], \qquad {\rm with} \qquad B^{IJ} = g^{\mu \nu} \partial_\mu X^I \partial_\nu X^J\, .$$ The energy momentum tensor for the fields $X^I$ is that of a perfect fluid with zero pressure, with the fluid trajectory given by the condition $X^I(x^\mu) = const.$~\cite{Dubovsky:2005xd,Endlich:2010hf}. It follows that in the corresponding classical system $X^I = const.$ are geodesics, as expected from the test-particle version~\eqref{2-intro}}.  However, as we are not aiming at solving any Wheeler-DeWitt equation in this paper, one can remain agnostic about the actual form of~\eqref{2}. What we really need to assume is that a theory like~\eqref{2} makes sense non-perturbatively and that the reference fields and the three-dimensional metric are good and commuting dynamical variables for the canonical quantization of the system,\footnote{No introduction nor self-contained review of canonical quantum gravity and ADM formalism will be attempted here. The unbeaten reference on the subject is still probably~\cite{DeWitt:1967yk}} so that a generic state will be a functional of these fields,
\begin{equation}  \label{portrayed}
\Psi = \Psi\left[h_{ij} (x^k), \, \phi(x^k), \, X^I(x^k), \, T(x^k)  \right]\, .
\end{equation}

\subsection{Low energy but non-perturbative}

It should be possible to capture  non-perturbative aspects of quantum gravity without know\-ledge of its UV-completion.  This is the broad point of view that motivates studying the black hole evaporation problem without relying on any specific UV-physics.
In theories with holographic duals the stage for non-perturbative low energy gravity has been usefully depicted in~\cite{Jafferis:2017tiu}. It is argued that, much like in a standard non-gravitational system, the long wavelength modes of the bulk theory are linearly projected into the total Hilbert space of the CFT. In other words, they still represent a subspace of the total Hilbert space.  Apparent contradictions and non-linearities related to the state dependence of bulk observables are taken care of by the standard diffeomorphism constraints of canonical gravity~\cite{Jafferis:2017tiu}, which restrict the physical Hilbert space and will be discussed momentarily. 

Inside the state $\Psi$ quantum fluctuations are expected to be  present up to the UV with an exponentially suppressed distribution. For example, the standard vacuum wavefunction of a free scalar  $\phi$ goes in Fourier space as 
\begin{equation}\label{vacuum}
\Psi[\phi_{\vec k}] \ \sim \ \exp\left(-\frac12\sum_{\vec k} k \ \phi_{\vec k} \phi_{-\vec k} \right).
\end{equation}
However, for the effective field theory to make sense, the low energy dynamics should be independent of such a ``UV tail". Following the reasoning of~\cite{Jafferis:2017tiu}, the short wavelength modes should be safely set in their vacuum state, and it should be possible to 
characterize a generic low-energy state by the long modes that are populated below the cutoff.   

With this caveat in mind,  it looks reasonable to make a ``low energy hypothesis" on the state $\Psi$,  
\begin{itemize}
\item {\bf Assumption 1}\ Scalar curvature invariant operators (both 3- and 4-dimensional) will acquire a whole range of values within $\Psi$, depending on the position at which they are evaluated and on the quantum fluctuations present in $\Psi$. We require that, well below the cutoff of the theory,  there is a \emph{maximal curvature scale} for such values, above which the corresponding probability rapidly drops to zero.
\end{itemize}

Next we should make sure that the fields $X^I$ and $T$ behave sufficiently well in the low energy theory so that they can be used as a reference. We want to consider highly quantum states, but within limits! For example, the gradients of the fields $X^I$ should be spacelike when calculated on the state $\Psi$, 
\begin{itemize}
\item
{\bf Assumption 2} \ Consider the three scalar operators $g^{\mu \nu} \partial_\mu X^I \partial_\nu X^I$ (no summation over $I$ here). We assume that \emph{all possible values that these operators can take when evaluated on $\Psi$ are strictly positive}---which in our mostly-positive convention for the metric means that the gradients are spacelike. Such gradients should also never become parallel to each other. To this purpose, we can introduce a ``vector product" operator parallel to the fluid velocity, 
\begin{equation}
J^u = \epsilon^{\mu \alpha \beta \gamma}\epsilon_{IJK} \partial_\alpha X^I \partial_\beta X^J \partial_\gamma X^K, 
\end{equation}
and ask that it be strictly non zero when evaluated on $\Psi$. 
\end{itemize}
\begin{itemize}
\item {\bf Assumption 3} \ The possible values that the operator
%\footnote{In the canonical formalism and for a canonical scalar field $T$, up to operator ordering issues,
%$$ g^{\mu \nu}\partial_\mu T \partial_\nu T = N\left(- \frac{ \hat \pi_T^2}{\sqrt{\hat h}} + \hat h^{ij} \partial_i \hat T \partial_j \hat T\right),$$
%where $\hat \pi_T$ is the conjugate momentum to $\hat T$ and $N$ is the lapse function.} 
$g^{\mu \nu}\partial_\mu T \partial_\nu T$ takes on the state $\Psi$ are \emph{strictly negative}. 
\end{itemize}  
 Again, these requirements can possibly make sense only if we can safely ignore the short wavelength modes, which are assumed to be in their vacuum. 
 
\subsection{Diffeomorphism constraints and states at ``equal time"} \label{sec_timefield}

We shall assume that the state~\eqref{portrayed} satisfies the momentum and Hamiltonian constraints of the theory. 
This means that $\Psi$ should evaluate the same on all different configurations of fields that are related by diffeormorphisms. Especially when time diffs are considered, this statement is notoriously tricky, in that the equivalence class of configurations also includes all possible different time-foliations of the system. 
Despite being a functional of \emph{spatial} field configurations, $\Psi$ really portrays an entire portion of the four-dimensional spacetime.

Thanks to the time field $T$ and to the strong assumption {\bf 3} above, we can extract from $\Psi$ the state of the system ``at a given time" by using $T$ as a clock. In fact, we can evaluate the state $\Psi$ on a section of the configuration space characterized by some constant value  of $T$, $T(x^k) = T_0$.  
\begin{equation} \label{psitilde}
\tilde \Psi[h_{ij} (x^k), \, \phi(x^k),\, X^I(x^k);T_0] \ \equiv\ {\cal N}(T_0) \, \Psi[h_{ij} (x^k), \, \phi(x^k), \, X^I(x^k),  \, T(x^k) =T_0] , \
\end{equation}
where ${\cal N}(T_0)$ is a normalization constant. This defines a conditional probability amplitude to find the remaining fields in some configuration \emph{given that} $T(x^k) = T_0$. Equivalently, this is the relative state \emph{\`a la} Everett~\cite{Everett:1957hd} for the rest of the system when the subsystem $T$ finds itself in the ``state $T_0$". If there is sufficient correlation, or entanglement, between $T$ and the  rest of the system, a smooth change of the value of $T_0$ will produce a continuum of different relative states $\tilde \Psi$, thus describing a ``time evolution" for the remaining fields. 
The relative state $\tilde \Psi$ is a picture of the system ``at a given time", made diff-invariant by the presence of the time field $T$. 

Note that conditional probabilities also allow to weaken assumptions {\bf 1}, {\bf 2} and {\bf 3}.  The latter do not need to be valid at the level of the entire state $\Psi$---which could contain e.g. the entire history of a closed universe with regions arbitrarily close to the singularity. For the use that we are going to make of them in what follows, assumptions {\bf 1}, {\bf 2} and {\bf 3} could as well just be valid within some time slice, i.e.,  subject to the condition that $T(x^k) = T_0$, for some range of values of $T_0$. 

The state $\tp$ depends on all fields except $T$ and is only subject to the momentum constraints. Again, this means that $\tp$ must evaluate the same on all field configurations that are related by \emph{spatial} diffeomorphisms. The restricted Hilbert space $\tilde {\cal H}$ to which $\tp$ belongs is only spanned by the functionals of the field configurations $h_{ij}(x^k)$, $\phi(x^k)$ and $X^I(x^k)$. Since the corresponding operators commute with each other, a convenient basis of $\tilde {\cal H}$ is made by the common eigenvectors of these operators, 
that we denote concisely by $|h,\phi, X^I\rangle $,
\begin{eqnarray}
\hat h_{ij}(x^k) |h,\phi, X^I\rangle = h_{ij}(x^k) |h,\phi, X^I\rangle , \quad
\hat X^I (x^k)|h,\phi, X^I\rangle = X^I (x^k)|h,\phi, X^I\rangle \, ,\ \ {\rm etc.}
\end{eqnarray}
On the LHS of these equations the ket is acted upon by the quantum operator while on the RHS it is simply multiplied by the corresponding eigenvalue. Of course, as usual, the wavefuntion $\tp$ defined in~\eqref{psitilde} is already the scalar product of the state $|\tp\rangle$ with the  generic element of this basis, i.e. 
\begin{equation} 
\tilde \Psi[h_{ij} (x^k), \, \phi(x^k),\, X^I(x^k);T_0] \ \equiv \ \langle h_{ij} (x^k), \, \phi(x^k),\, X^I(x^k) | \tp;T_0\rangle\, ,
\end{equation}
where a more explicit notation for the basis vectors has been used for clarity. 

The meaning of the states $|h,\phi, X_I\rangle $ is quite straightforward. We call these vectors a \emph{classical basis} because they represent classical \emph{spatial} geometries with perfectly 
defined field configurations and localized observers. Note however that these states have infinite uncertainties on the conjugate momenta, and are thus very bad approximations of classical space\emph{times}. 

Because of momentum constraints, the classical basis $|h,\phi, X^I\rangle $ is overcomplete. 
Infinitely many  classical field configurations related by spatial diffeomorphisms correspond to the same physical state.  For each class of gauge-equivalent metrics we should thus pick a representative, that we denote by $\tilde h$ and restrict our functional integrals accordingly
\begin{equation}
{\cal D} h \longrightarrow {\cal D} \tilde h .
\end{equation}
In the case of a closed universe, $\tilde h$ really represents some three-dimensional geometry, rather than a tensor field~\cite{DeWitt:1967yk}. When taking quantum averages, we should sum over the restricted set of tensors $\tilde h$ and not over $h$. 
So a resolution of the identity for the restricted Hilbert space $\tilde {\cal H}$ is written as
\begin{equation}
\mathbb{1}\ = \ \sum_I\int {\cal D} \tilde h\,  {\cal D} \phi\,  {\cal D} X^I\, \    |h,\phi, X^I\rangle  \langle \tilde h,\phi, X^I|\, .
\end{equation}

\subsection{Distance operator} \label{sec_distance}

We can now attempt a formal definition of the distance operator. 
As said, a triplet of values for the fields $X^I$ identifies a given observer. However, on the eigenbasis $|h,\phi, X^I\rangle $, they also entirely define a point in space. Let us indicate such triplets of values by $\vec X$, $\vec Y$ etc. For example, we could consider $\vec X = \{1,0,0\}$, $\vec Y = \{0,1,0\}$. 
The quantum operator $\hat d(\vec X, \vec Y)$ ``distance between observer $\vec X$ and observer $\vec Y$" (what would correspond in the discrete model~\eqref{2-intro} to a distance operator between, say, observer $A$ and observer $B$) is straightforwardly defined on this basis, although in the usual implicit manner. One has to find the spatial geodesic between the corresponding points for the geometry $h_{ij}(x^k)$ and then calculate its length by integrating the differential line element. The outcome of this operation has no uncertainty on the classical basis and thus $|h,\phi, X^I\rangle $ must be an eigenvector of $\hat d(\vec X, \vec Y)$,

\begin{equation}
\hat d(\vec X, \vec Y) |h,\phi, X^I\rangle =  d(\vec X, \vec Y)_{\{h,  X^I\}} |h,\phi, X^I\rangle\, .
\end{equation}
The  notation above wants to emphasize that the eigenvalue of the distance operator depends only on the geometry and on the position of the observers, not on the other fields.

One can formally express the expectation value of any power of the distance operator on the state $\tilde \Psi$ as 
\begin{equation}
\left\langle  \hat d(\vec X, \vec Y)^n  \right  \rangle =  \sum_I \int {\cal D} \tilde h\,  {\cal D} \phi\,  {\cal D} X^I 
\left|\langle \tilde  \Psi | h,\phi, X^I\rangle\right|^2 \ \left(d(\vec X, \vec Y)_{\{ h,  X^I\}}\right)^n \, .
 \end{equation}
One could simply consider the average of the distance operator $\langle \hat d \rangle$ and find that it has interesting non-additive properties. However this quantity does not have a very smooth local limit. 
The expression that best generalizes standard classical distances, we argue, is what we call in short \emph{beyond-Riemannian distance} (\emph{BR-distance}), defined as
\begin{equation}
\overline{d(\vec X,\vec Y)} \equiv \sqrt{\left\langle \hat d(\vec X,\vec Y)^2\right\rangle}\, .
\end{equation}

Our formulas make implicit use of the standard ``\emph{squared-norm}" scalar product. In quantum gravity one often adopts the Klein-Gordon product among states, because the latter is guaranteed to be invariant under time diffeomorphisms~\cite{DeWitt:1967yk}. However, as already noted, the vectors $\tp$ in the restricted Hilbert space $\tilde {\cal H}$ represent states at a given time and do not need to satisfy the Wheeler-DeWitt equation. So it seems consistent, albeit not necessarily correct, to use the standard  quantum mechanics product in $\tilde {\cal H}$.
 
 \section{Properties of distances on quantum states}\label{sec_avdis}
 
In this section we explore the properties of the expectation value defined above and show that it can be different than the corresponding classical quantity in a standard three-dimensional manifold. 
In order to perceive deviations from the classical behavior we need to consider ``highly quantum states" $\tilde \Psi$ on which the fluctuations of the distance operator are relevant. 

The state $\tp$ clearly contains possible uncertainties in both the metric field and the observers' positions. 
One naive but legitimate question is then the following. Can we distinguish between $(i)$ \emph{states where the metric is highly fluctuating} from $(ii)$ \emph{states where the positions of the observers are highly fluctuating}? As we are going to see, there is no clear separation between these two situations, which reiterates the known statement that observables are relational in quantum gravity. However, one of the two limiting situations $(ii)$ is clearly well defined. Imagine that  $\tilde\Psi$ is extremely peaked on some geometry $\tilde h$ but highly uncertain about the observers' configuration $X^I(x^k)$. This would clearly represent situation $(ii)$, where the indeterminacy of the distance operator is entirely due to the spreading of the observers' positions. The opposite situation, on the other hand, does not appear to be well defined. In other words, there seems to be no sense in which the metric fluctuates but the observers are kept ``fixed". 

Let us look more deeply into these issues with the aid of specific coordinate choices.

\subsection{Unitary gauge}\label{sec_unitgauge}

The reference fields $X^I$ and $T$  allow to understand diffeomorphism constraints in a different, perhaps more mundane, perspective. In the semiclassical treatment one can directly write an action like~\eqref{2} in \emph{unitary gauge}, where the coordinates are fixed to coincide with the reference fields. The latter effectively disappear as dynamical variables and the corresponding degrees of freedom are ``eaten" by the metric (see e.g.~\cite{Piazza:2013coa} and references therein). In the canonical formalism 
the values of the lapse and the shift are no longer arbitrary and must be expressed in terms of the remaining fields by solving the constraint equations. 
Explicit diffeomorphism invariance is lost  in the process, which results, for instance, in having terms in the gauged-fixed action where the spatial metric appears without derivatives. 

Using the inflaton as a clock is by now standard practice in inflationary cosmology (e.g.~\cite{Maldacena:2002vr,Cheung:2007st}). 
Of course, at the full non-perturbative level, this gauge-fixing process goes through a number of caveats and subtleties to which we can only oppose fierce and imprudent optimism. However, by quantizing the system directly in the unitary gauge for the time field $T$, one would end up with a wavefunction which depends only on the remaining fields, i.e., the reduced wavefunction $\tp$.

But one can go one step further and fix also 
the spatial coordinates $x^i$ in such a way that they coincide with the values of the fields $X^I$, 
\begin{equation}
 X^I = x^i \ ; \qquad \frac{\partial X^I}{\partial x^i} = \delta^I_i  .
\end{equation}
 The degrees of freedom previously contained in the fields $X^I$ have migrated into the metric.  
 In this gauge the wavefunction depends only on the remaining variable, 
 \begin{equation}
\tilde \Psi_{\rm U} = \tilde \Psi_{\rm U}[h_{IJ} (X^K), \, \phi(X^K);T_0] \, ,
\end{equation}
(the pedix U stands for unitary gauge). These coordinates have physical meaning because they are attached to physical observers. The momentum constraint is automatically satisfied in this gauge. We can label the vectors of the classical basis with the values that the metric and the remaining fields assume in unitary gauge, 
\begin{equation}
|h_{IJ} (X^K), \, \phi(X^K)\rangle_{\rm U}\, .
\end{equation}
The above basis is not overcomplete, and we do not need to restrict the integration measure any longer in the functional integrals (see e.g. eq.~\eqref{averageUG} below). 
The reason is that two metrics $h_{IJ}(X^I)$ and $h'_{IJ}(X^I)$ which differ by a  coordinate transformation still represent the same geometry but with the observers distributed differently. 
Therefore, they represent two different states. Equivalently, the two corresponding configurations can give 
distinct probability amplitudes when evaluated within~$\tilde \Psi_{\rm U}$.

In this gauge the position of the observers does not seem to fluctuate. However, the situation $(ii)$ depicted at the beginning of this section---i.e. ``\emph{only the observers fluctuate}"---is easily recognizable. 
This will be some $\tilde \Psi$ whose entire support is on a set of metrics $h_{ij}$ that are all equivalent  to one single metric up to gauge transformations. 

Let's make an example also for later use. On a two-dimensional sphere we smoothly rotate the two poles with respect to the equator of an angle $\epsilon$. This is achieved by a coordinate transformation
\begin{equation}\label{transtrans}
\Theta \rightarrow  \Theta\, , \qquad \varPhi \rightarrow  \varPhi + \epsilon f(\Theta)\, ,
\end{equation}
\begin{figure}[h]
\captionsetup{width=.9\linewidth}
\begin{center}
\includegraphics[width=6cm]{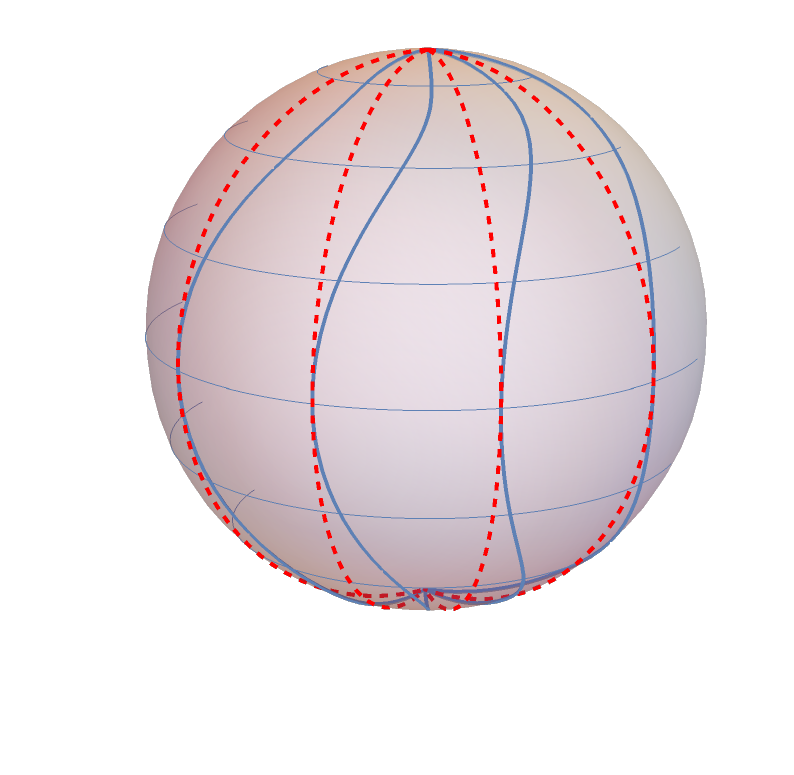}
\end{center}
\caption{On the same geometry (a sphere), red dashed and thick blue lines represent $\varPhi = const.$ observers for $\epsilon = 0$ and $\epsilon = 0.4$ respectively. The choice of function here is $f(\Theta) = 1 - e^{-\tan^2(\Theta - \pi/2)}$  \label{figobs}}
\end{figure}
where $\Theta \in[0,\pi]$ and $\varPhi\in[0,2 \pi]$  and the smooth function $f(\Theta)$ is such that $f(0) = f(\pi) = 1$ and has a minimum at $f(\pi/2) = 0$. For instance, $f(\Theta) = 1 - e^{-\tan^2(\Theta - \pi/2)}$ will do the job (see Fig.~\ref{figobs}).  
The metric in the new coordinates reads 
\begin{equation}h_{IJ}^{(\epsilon)} \ = \ \begin{pmatrix} 1+ \epsilon^2 {f'(\Theta)}^2 \sin^2 \Theta & - \epsilon f'(\Theta) \sin^2 \Theta  \\
- \epsilon f'(\Theta) \sin^2 \Theta & \sin^2 \Theta\, 
\end{pmatrix}\, .
\end{equation}
We stress again that $\Theta$ and $\varPhi$ are the labels, the ``names", of our distinguishable observers and thus different values of the parameter $\epsilon$ represent different states. 
Say that $|h^{(\epsilon)}(\Theta,\varPhi)\rangle_{\rm U}$ is the corresponding eigenvector in the presence of some fixed matter fields' configuration\footnote{We will pay no attention to the remaining fields $\phi$ from now on and thus we will omit them from the notation}. 
The state 
\begin{equation}\label{smoother}
|\tilde \Psi\rangle  = \int d\epsilon \ \psi(\epsilon) |h^{(\epsilon)}(\Theta,\varPhi)\rangle_{\rm U} ,
\end{equation}
where $\psi(\epsilon)$ is some suitable probability amplitude, is entirely supported on metrics that represent the same geometry. We can thus say that the uncertainty of the corresponding distances in this state is due to the uncertainty in the (relative) positions of the observers. 

As already mentioned, however, there is no such thing as situation $(i)$ in this framework, where ``\emph{only the metric fluctuates}". In this gauge one could try to build a state $\tilde \Psi$ that has support on a continuous class of metrics that are all ``transverse" to one another, so that they are not reachable one from another by means of a gauge transformation. However, this automatically implies also a relational displacement among our observers. The point is that even the value of the metric tensor in a point is meaningful in this gauge, because it tells the distances between the observers that are infinitesimally close to that point. In order to see this in more detail, let us discuss another choice of coordinates. \emph{From now on we omit the matter fields for notational convenience.}

\subsection{Normal coordinates}\label{sec_normal}

Let us focus on the observer who sits at the origin of the coordinate system in unitary gauge, $X^I=0$. Let us decompose the state $\tilde \Psi$ into the eigenvectors of the metric field $|h_{IJ}(X^K)\rangle_{\rm U}$. Every such state represents a definite classical metric, which we can express in Riemann normal coordinates around our chosen observer, 
\begin{equation} \label{riemnor}
h_{ij} (y^k) = \delta_{ij} - \frac13 R_{ikjl} \ y^k y^l + \dots\, ,
\end{equation}
where we have set $y^k = 0$ to coincide with $X^I = 0$. For any metric~$h_{IJ}(X^I)$ on which $\tilde \Psi_{\rm U}$ has support, there is a transformation law connecting Riemann normal coordinates and unitary gauge coordinates, 
\begin{equation} \label{transformation}
y^k = y^k (X^I) \simeq A^k_{\ I}\, X^I + B^k_{\ IJ}\, X^I X^J + \dots\, .
\end{equation} 
The coefficients $A$, $B$ etc. of this expansion depend on the functional form of the metric in a neighborhood of the origin. In particular, $A^k_{\ I}$ depends just on the actual value $h_{IJ}(0)$,  $A^k_{\ I} A^k_{\ J} = h_{IJ}(0)$.  This transformation is defined up to a global rotation $R$ (e.g. $A^k_{\ I} \rightarrow R^{kl} A^l_{\ I}$ etc.) that leaves the leading flat term in the metric~\eqref{riemnor} invariant. Before moving further it is convenient to get rid of this ambiguity and fix a convention for such a rotation. At linear order in the transformation~\eqref{transformation} we can ask the coordinate $y^1$ to be parallel to $X^1$. We are then left to fix the rotation around such a common axis, for instance by insisting that the axis $y^2$ belongs to the $X^1 \!-\! X^2$ plane. 
All this corresponds to choosing
\begin{equation}
A^1_{\ 2} = A^1_{\ 3} = A^2_{\ 3} = 0\, .
\end{equation}

By inverting~\eqref{riemnor} we can then express $X^I$ in normal coordinates. Our ``classical basis" of vectors which in unitary gauge are specified by the metric configuration $h_{IJ}(X^K)$, are now defined by the curvatures in the origin and by the positions of the observers around the origin,
\begin{equation}\label{clasnc}
|h_{IJ}(X^I)\rangle_{\rm U}\ \simeq  \ |X^I(y^k), R_{ikjl}(0), \dots\rangle_{\rm NC}\, ,
\end{equation}
where the ellipsis stands for the tower of higher order curvature tensors needed to expand the metric to an arbitrary order.  By superposition,  a generic state is expressed in this gauge as a functional of the observers' positions and a function of the curvatures in the origin,
 \begin{equation}
\tilde \Psi_{\rm NC} = \tilde \Psi_{\rm NC}[X^I(y^k), R_{ikjl}(0), \dots;T_0] \, .
\end{equation}
The normal coordinate (NC) gauge conveniently conveys the observer's viewpoint. We want to discuss in turn the dependence of this wavefunction on the curvatures in the origin and on the observers positions. 

\subsection{Immediate neighborhood: local flatness} \label{2.1}
 
In the previous section we have formulated a ``low energy hypothesis" on the state $\tp$, assumption {\bf 1}. Again, if we want to make sense of the low energy theory we should assume an upper bound on all possible value that the curvature invariants can take when calculated on $\tp$. 
If this hypothesis applies, the NC-gauge perspective suggests that \emph{there is always a neighborhood around each observer that looks locally flat.} This ``quantum version" of the equivalence principle is suggested by the fact that 
being locally flat around one point is nothing more than a gauge condition. The NC gauge is not necessarily convenient to quantize the system~\cite{Nelson:1988ta} but it explicitly shows this point. Even if the gravitational wavefunction is supported on a range of widely different metrics, all such metrics look flat around each observer  in this gauge, if we are granted an upper bound on the possible values of the \emph{curvature}.  The one metric in the quantum ensemble with the highest curvature will set the length within which the equivalence principle applies. 

Within the locally flat neighborhood of each observer we can still have a large uncertainty in the positions of the nearby observers.   Let us see this in more detail.

On each element of the classical basis~\eqref{clasnc} the distance from the origin is simply the Euclidean distance expressed in the $y^k$ coordinates. 
This means that, on such basis, 
\begin{align}
\hat d(0, \vec Y)^2 |X^I(y^k), \dots\rangle_{\rm NC}\ &= \ y^k(\vec Y) \ y^k(\vec Y) \ \, |X^I(y^k), \dots\rangle_{\rm NC} \\[2mm]
& = \ A^k_{\ I} A^k_{\ J} Y^I Y^J \ |X^I(y^k), \dots\rangle_{\rm NC} + {\cal O}(Y)^3\, , 
\end{align}
where in the last line the matrix $A^k_{\ I}$ is the linear piece of the coordinate transformation~\eqref{transformation} relative to  the configuration $X^I(y^k)$ of that basis vector. In other words, $X^I(y^k)$ can be considered as a proxy for the matrices $A^k_{\ I}$, $B^k_{\ IJ}$ etc. that perform the transformation $y^k \rightarrow X^K$. This observation justifies the change of variable in the following functional integral, where we express the average of the square distance of a nearby observer, 
\begin{align}
\left\langle \hat d(0, \vec Y)^2 \right\rangle\ &\simeq\ \int {\cal D} [X^I] \left|\tp_{\rm NC}[X^I(y^k)]\right|^2 A^k_{\ I} A^k_{\ J} Y^I Y^J \label{4.14}\\
&\simeq \ \int  {\cal D} [(A^{-1})^{\ I}_k] \left| \tp_{\rm NC}[(A^{-1})^{\ I}_k]\right|^2 h_{IJ}(0)\  Y^I Y^J \label{4.15} \\ \label{4.16}
&= \ {\cal G}_{IJ}  \ Y^I Y^J\, ,
\end{align}
where $A^k_{\ I} A^k_{\ J} = h_{IJ}(0)$ has been used in going from~\eqref{4.14} to~\eqref{4.15}. So in the vicinity of each point the BR-distance of the nearby observers can be smoothly expressed by the effective metric ${\cal G}_{IJ}$ which is an appropriate average of all metrics $h_{IJ}(0)$,  
\begin{equation} \label{EP}
\overline{d(0,\vec Y)} \ =  \ \sqrt{{\cal G}_{IJ}  \ Y^I Y^J }\ + \ {\cal O}(Y)^{3/2}\, .
\end{equation}
This is the claimed local limit of BR-distances.
 
The uncertainty in the unitary gauge metric $h_{IJ}(0)$ thus translates, in NC-gauge, into the uncertainty in the distances of the nearby observers. This uncertainty grows with the distance itself and,  in any reasonable scenario, we shold assume that it is extremely small.  So that when calculated, say, at the typical curvature scale, it represents a small fraction of the curvature scale itself. 

If the fluctuations of the metric vanish at one point, then the uncertainty of the nearby observers' positions also vanishes at linear order. This is what happens, for dynamical reasons, at the boundary of asymptotically AdS. In this context a nice analogue NC gauge are Fefferman-Graham coordinates, as we detail in the Appendix.

\subsection{A little further away: average distances are non-additive}

As previously discussed, the average distance between two observers is an implicit but rather well-defined calculation in this framework. Let us summarize here the procedure by specializing the previous formulae to the unitary gauge and by omitting the matter fields for notational convenience. 

Our classical basis of vectors in unitary gauge is $|h_{IJ}(X^I)\rangle_{\rm U}$. They represent classical field configurations with metric and observers entirely defined. They are therefore eigenvectors of the distance operator, 
\begin{equation}
\hat d(\vec X, \vec Y) |h_{IJ}(X^I)\rangle_{\rm U} =  d(\vec X, \vec Y)_{\{h\}} |h_{IJ}(X^I)\rangle_{\rm U} .
\end{equation}
The eigenvalue $d(\vec X, \vec Y)_{\{h\}} $ depends on the metric because given some metric one can find the geodesic connecting $\vec X$ and $\vec Y$ and then calculate its length. When the average of some power of the distance operator is calculated on some generic vector, 
\begin{equation} \label{averageUG}
\left \langle  \hat d(\vec X, \vec Y)^n   \right \rangle =  \int {\cal D} h\,   
\left|\langle \tilde  \Psi |h_{IJ}(X^I)\rangle_{\rm U}\right|^2 \ \left(d(\vec X, \vec Y)_{\{h\}}\right
)^n \, ,
 \end{equation}
this procedure should be done for every element of the basis on which $|\tp\rangle$ decomposes. 

Something interesting happens if geodesics calculated for (slightly) different basis vectors $|h_{IJ}(X^I)\rangle_{\rm U}$ inside the ensemble take (very) different paths. Let us consider the following one-dimensional example. 

\subsubsection*{Example 1}
The state $\tp$ is a superposition of only two vectors belonging to the ``classical basis",
\begin{equation}
|\tp \rangle = \frac{1}{\sqrt{2}}\left(|\tp_1 \rangle + |\tp_2 \rangle\right)
\end{equation}
where the classical states $|\tp_1 \rangle$ and  $|\tp_2 \rangle$ are depicted in the figure below. Such a superposition gives the following expectation values for the distances:
\begin{figure}[h]
  \centering %\vspace{-.2cm}
  \includegraphics[width = .9\linewidth]{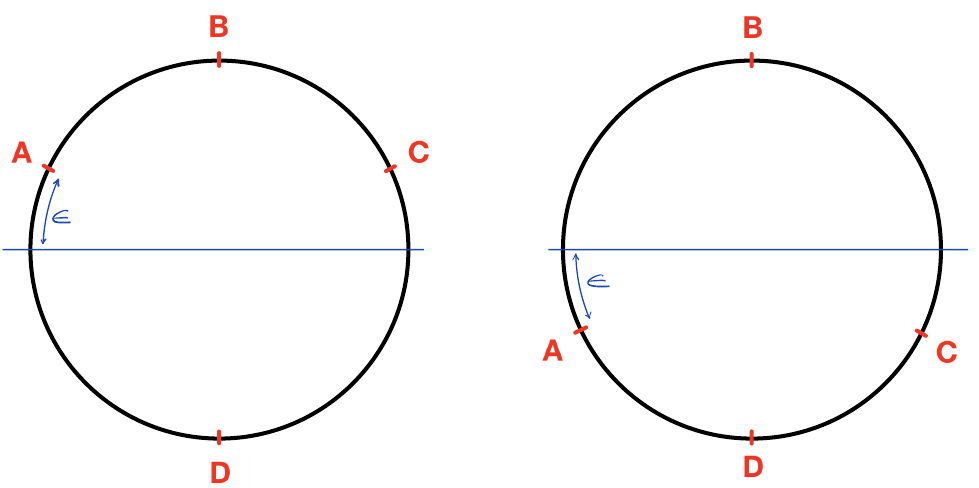} \\[.2cm]
   \Large $\displaystyle \ |\tilde \Psi_{1} \rangle  \ \ \qquad \quad \qquad \qquad \qquad  \qquad \qquad |\tilde \Psi_{2} \rangle   $
  \end{figure}
\begin{align}
\langle \hat d(A,B) \rangle & = \langle \hat d(B,C) \rangle = \langle \hat d(A,D) \rangle = \langle \hat d(C,D) \rangle = \ell \, ,\\
\langle \hat d(B,D) \rangle & = 2 \ell \, ,\\
\langle \hat d(A,C) \rangle & = 2 (\ell - \epsilon) \, ,
\end{align}
where $\ell = r \pi /2$, $r$ being the radius of the circle. To first order in $\epsilon$ these are also the BR-distances between the observers $\overline{d(A,B)} \simeq \langle \hat d(A,B) \rangle$ etc. No Riemannian one-dimensional manifold can reproduce the above set of distances. 
The point is simply that \emph{quantum averaged} Riemannian distances behave differently than Riemannian distances. This anomaly is not due  to the topology of the problem (the fact that the circle is not simply connected) as our second example shows.

\subsubsection*{Example 2}
Let us go back to the example of Sec.~\ref{sec_unitgauge} and chose the state as follows
\begin{equation}
|\tp \rangle = \frac{1}{\sqrt{2}}\left(|h^{(\epsilon)}(\Theta,\varPhi)\rangle_{\rm U} + |h^{(-\epsilon)}(\Theta,\varPhi)\rangle_{\rm U} \right)\, 
\end{equation}
(any smoother superposition would also do the job). We can evaluate the eigenvalue of the distance operator on each classical state $|h^{(\epsilon)}(\Theta,\varPhi)\rangle_{\rm U}$ by using the ``haversine formula" for distances on a sphere of unit radius. In standard coordinates it reads
\begin{equation}
d\left(\{\theta_1,\varphi_1\},\{\theta_2, \varphi_2\}\right)_{\{\epsilon = 0\}} = 2 \arcsin\sqrt{\sin^2\frac{\theta_1-\theta_2}{2} \ +\ \sin(\theta_1)\sin(\theta_2) \sin^2\frac{\varphi_1-\varphi_2}{2}}.
\end{equation}
Then we should perform the coordinate transformation~\eqref{transtrans},
\begin{align}
\Theta & = \theta\\
\varPhi & =  \varphi + \epsilon f(\theta) \\ 
& = \varphi + \epsilon \left(1 - e^{-\tan^2(\theta - \pi/2)}\right)\, ,
\end{align}
where a specific choice for $f$ has been made, in order to obtain $d(\{\Theta_1,\varPhi_1\},\{\Theta_2, \varPhi_2\})_{\{\epsilon\}}$.\footnote{Notice that in this two-dimensional example we are pedantically indicating the observer $\vec X$ with a doublet of components  $\{\Theta , \varPhi\}$.}
The outcome is a rather complicate expression 
 which however can be simplified in some cases. 
\begin{figure}[h]
\captionsetup{width=.9\linewidth}
\begin{center}
\includegraphics[width=8cm]{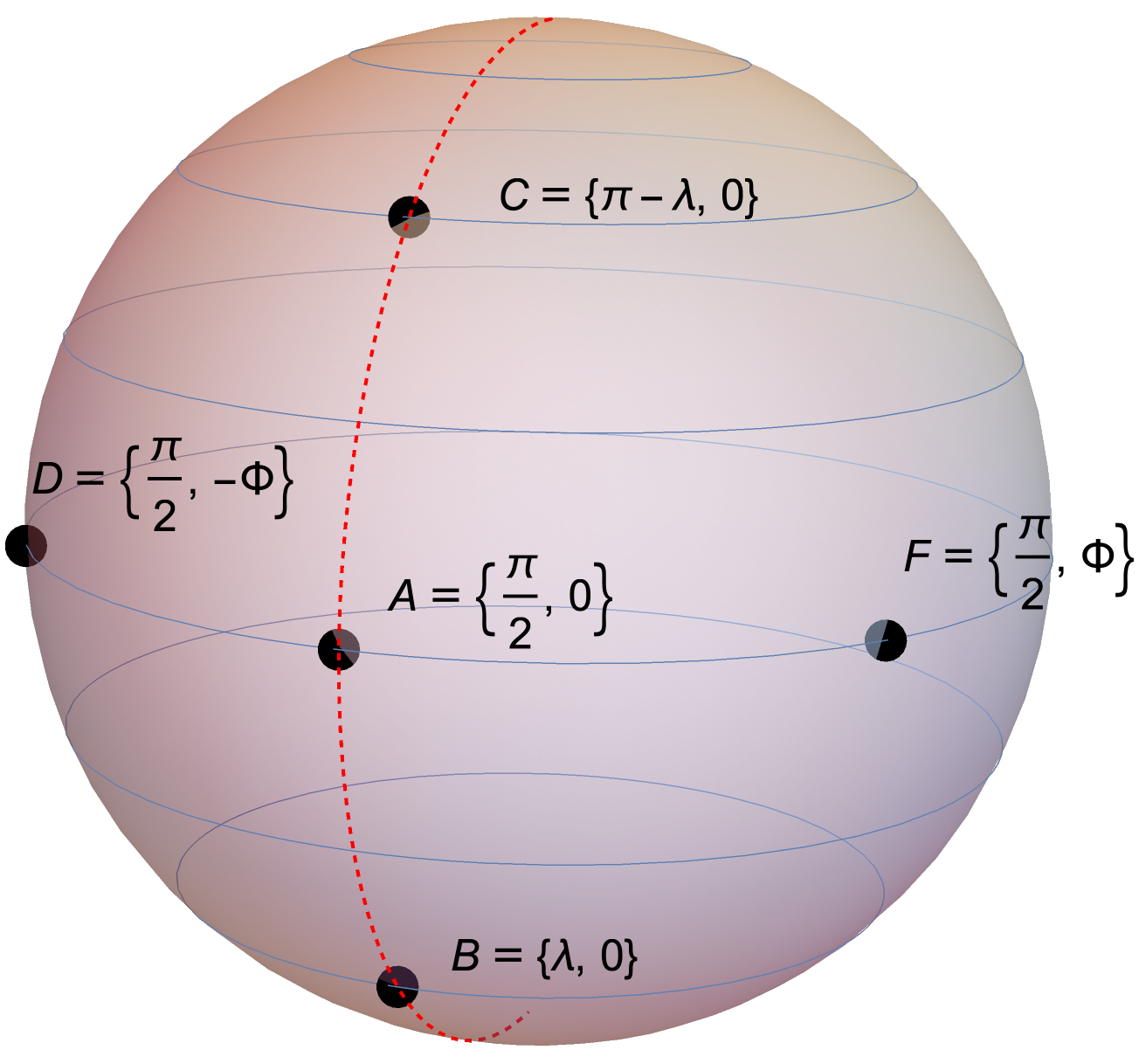}
\end{center}
\caption{A Lorentzian manifold embedded into a higher dimensional one. The intrinsic lightcone structure is represented by dashed blue lines. Null geodesics in the ambient space (the red line) can connect events that are at spacelike separation.\label{spBR}}
\end{figure}
\begin{enumerate}
\item The distance between two points at the same longitude, say $\varPhi_1 = \varPhi_2 = 0$ and of opposite latitude (i.e. one at distance $\lambda$ from the north pole and the other at the same distance $\lambda$ from the south pole), $\Theta_1 = \lambda, \Theta_2 = \pi - \lambda$, is simply
\begin{equation}\label{limit1}
d(\{\lambda,0\},\{\pi-\lambda, 0\})_{\{\epsilon\}} = \pi - 2\lambda\, .
\end{equation}
The result is independent of $\epsilon$ because of the parity of the function $f$ around the equator. Geodesics between points at the same longitude simply follow a meridian.

\item Let us now consider instead the distance between one of the two points above, $\{\lambda,0\}$, and a point on the equator at some longitude $\varPhi$, $\{\pi/2, \varPhi\}$. Such a distance should now depend on $\epsilon$ because the point near the pole ``has rotated" while the point on the equator has not. If the first point is very close to the pole we can assume that the function $f$ is very close to one and the relative rotation between the pole and the equator is $\epsilon$. In this limit we have
\begin{equation} \label{limit2}
d(\{\lambda,0\},\{\pi/2, \varPhi\})_{\{\epsilon\}} = \frac{\pi}{2} - \lambda \cos(\varPhi + \epsilon) + {\cal O}(\lambda^2)\, .
\end{equation}
\end{enumerate}
We can now focus on 5 specific observers, $A,B,C,D,F$ as defined and displayed in Fig.~\ref{spBR}.
We have all the instruments to calculate the average squared distances among them. For instance, 
\begin{equation}
\left \langle \hat d(A,B)^2 \right \rangle = \frac12\left(d(A,B)^2_{\{+\epsilon\}} + d(A,B)^2_{\{-\epsilon\}}\right) \quad {\rm etc.}\, ,
\end{equation}
as follows from the general formula~\eqref{averageUG}. Let us now compute BR-distances $\overline{d(A,B)} = \sqrt{ \langle \hat d(A,B)^2\rangle}$ etc. By using~\eqref{limit1} and~\eqref{limit2} we obtain
\begin{align}
\overline{d(A,B)} & = \overline{d(A,C)} = \frac{\pi}{2} - \lambda \cos \epsilon + {\cal O}(\lambda^2)\, ,\\
\overline{d(A,D)} & = \overline{ d(A,F)} = \varPhi \, , \quad  \overline{ d(D,F)}  = 2 \Phi \\
\overline{ d(D,C)} & = \overline{ d(C,F)} = \overline{ d(D,B)}  = \overline{ d(B,F)} = \frac{\pi}{2} - \lambda \cos \varPhi \, \cos \epsilon  + {\cal O}(\lambda^2) \, , \\
\overline{ d(B,C)} & = \pi - 2 \lambda \, .
\end{align}

Again, the direct average distances $\langle \hat d(A,B)\rangle$ would give the same results at this order in $\lambda$. 
The most striking aspect of this set of distances is their non-additivity  along the meridian, 
\begin{equation}
\overline{d(A,B)} + \overline{d(A,C)} > \overline{d(B,C)}
\end{equation}
for $\epsilon \neq 0$. 
The point is that the geodesics between $B$ and $C$ do not pass through $A$ for any of the two states in the classical basis, they cut through a shorter path. The observers $D$ and $F$ are ``control observers", symmetrical with respect to the meridian. Distances along the equator are not anomalous in this example. 

In summary, the observers in this example describe a two dimensional manifold that we can label with the coordinates $\Theta$ and $\varPhi$. There is a distance naturally defined on this manifold,  $\overline{d}$. Equipped with such a distance, the manifold is a metric space. Such a distance resembles a standard Riemannian distance at short separations. However it behaves anomalously at larger separations. In particular it looses the additivity property that Riemannian distances enjoy along any geodesic. It is a \emph{beyond-Riemannian} distance. 

One could object that these examples are too stupid. After all they deal with superpositions of identical geometries on top of which we are just moving the observers around. As already emphasized, however, the basis states that we are superimposing are two honestly different states of the theory. Moreover, fluctuations in the geometry would always imply also fluctuations in the relative positions of the observers as we have seen in Sec.~\ref{sec_normal}. At any rate, it would be easy to slightly perturb the two states that we are superimposing in such a way that they would also differ in their geometries, and 
still maintain the results above.

\section{Sketches of beyond-Riemannian geometry}\label{sec_BR}

\begin{figure}[h]
\centering
\captionsetup{width=.9\linewidth}
\subfigure[]{
\includegraphics[width=0.4\textwidth]{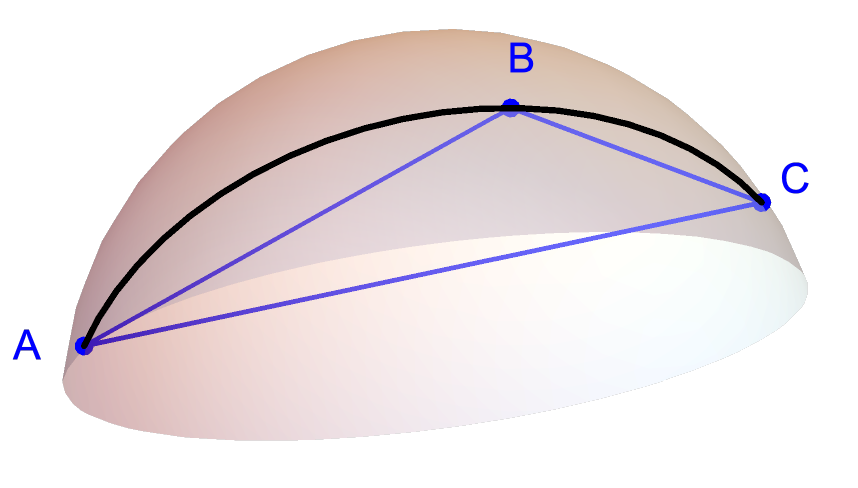}
\label{subfig:sphere}
}
\hspace{2cm}
\subfigure[]{
\includegraphics[width=0.4\textwidth]{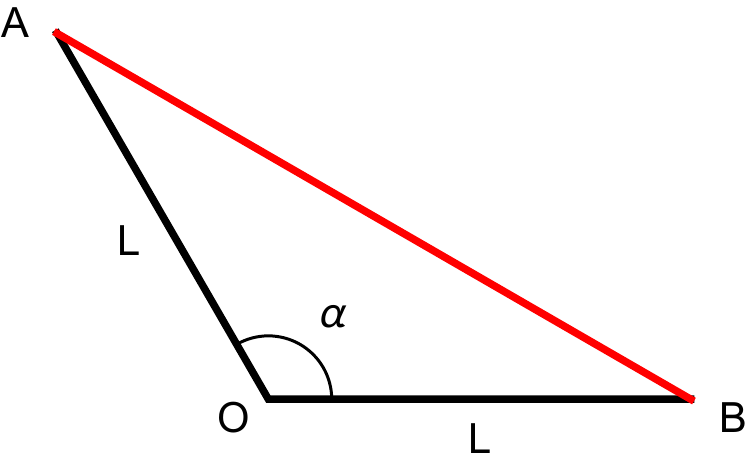}
\label{subfig:triangle}
}\\
\caption{\subref{subfig:sphere} Chord distances (in blue), as opposed to arc distances (in black), are non-additive. Although $A$, $B$ and $C$ are aligned on the sphere, $d(A,B)+ d(B,C)\neq d(A,C)$. \subref{subfig:triangle} What is the length of $AB$ given $L$ and the angle $\alpha$? With this triangle diagnostic we measure discrepancies from flat space induced by Riemannian and beyond Riemannian geometry. In both cases the corrections are of order $L^2$ with respect to flat space. However, the dependence on the angle $\alpha$ is different in the two cases [eq.~\eqref{1} vs. eq.~\eqref{3}] \label{fig:1}  }
\label{fig:1}
\end{figure}

We now attempt to sketch the general geometrical properties of BR-distances. A more
throughout study of this geometry will be given elsewhere~\cite{CPRZ}.
The geometrical description that we are after has to recover some notion of ``local flatness" around each point.  A natural generalization of Riemannian distances with this feature is suggested by \emph{chord} distances in embedded manifolds. Consider a two dimensional sphere embedded in the standard three dimensional euclidean space as in Fig.~\ref{subfig:sphere} and say that we define distances on the sphere by ``cutting through" the ambient space.  This distance is not additive and thus cannot be obtained as an integral of a line element. Nevertheless, such a metric space has the desired locally flat limit to satisfy the equivalence principle---eq.~\eqref{EP}, more precisely. Corrections to flat space geometry are of the same order as the standard Riemannian ones (${\cal O}(d^2/r^2)$ where $d$ is the size of the region considered and $r $ the radius of the sphere), but qualitatively different. Most notably, three aligned points (along the same geodesic) fail to saturate the triangle inequality because chord distances are not additive.

\subsection{The triangle diagnostic} \label{2.2}

Of course the choice of distances on a manifold can be highly arbitrary. Nevertheless their local behavior depends on a finite number of parameters. In order to better understand this we can use triangles as a diagnostic while taking the limit for the size of the triangle that goes to zero.  From some point $O$ we send two geodesics of equal length $L$ to the two points $A$ and $B$ as in Fig.~\ref{subfig:triangle}. The two curves span an angle $\alpha$ between them. 
In a flat Euclidean space the distance between $A$ and $B$ is  $d_{\rm Euclidean} = L\sqrt{2(1-\cos\alpha)}$.
\subsubsection{Standard-Riemannian discrepancies}
But in the presence of curvature the distance gets corrected by a fractional amount
\begin{equation} \label{1}
\delta_R = - \frac{R\, L^2}{24}(1+\cos \alpha) + {\cal O}(L^4)\, , 
\end{equation}
where $R$ is the Ricci scalar of the two dimensional surface subtended by the two vectors $OA$ and $OB$ and we have defined
\begin{equation} \label{discrepancy}
\delta_R = \frac{d_{\rm Riemannian} - d_{\rm Euclidean}}{d_{\rm Euclidean}}\, .
\end{equation}
As noted, the Riemannian discrepancies vanish at $\alpha = \pi$, when the three points $A$, $O$ and $B$ are aligned. This is when the problem becomes one-dimensional, and one-dimensional manifolds have no intrinsic curvature. Whenever we reiterate that ``Riemannian distances are additive" in this paper we mean, more precisely, that $\delta_R =0$ when $\alpha = \pi$. 

Another feature of standard Riemannian discrepancies is isotropy on any given plane. Once the plane subtended by the two vectors $OA$ and $OB$ is assigned within a manifold of arbitrary dimension, the result in~\eqref{1} only depends on the angle $\alpha$ between them and not on their overall orientation. 

\subsection{Beyond-Riemannian discrepancy}
Let us now look at what type of discrepancies we can expect in a beyond-Riemannian scenario. The triangle construction can still be made rigorously in this extended framework but the details will be given elsewhere~\cite{CPRZ}. For now let us look at the concrete example of embedded manifolds for inspiration and discuss the corrections induced by chord distances. Locally, they can be expressed by means of the extrinsic curvature of the embedding, which potentially introduces a lot of new degrees of freedom. Such corrections are generally anisotropic, in the sense that they depend on the orientation of $AB$ with respect to the principal axes of the extrinsic curvature. For example, if we embed our manifold ``like a cylinder", the correction vanishes when $AB$ is parallel to the axis of the cylinder and extremizes when it is orthogonal.
Among all such possibilities it looks natural to restrict to the isotropic case and assign to beyond-Riemannian corrections the same number degrees of freedom as the Riemannian ones. This corresponds to a ``sphere-like" embedding where the extrinsic curvature tensor is proportional to the identity, i.e. $K^i_j = \delta^i_j/r$ for some embedding radius $r$ which possibly depends on the plane to which the two vectors $OA$ and $OB$ belong. It is not difficult to see that this prescription leads to a beyond Riemannian discrepancy with respect to flat space of the form 
\begin{equation} \label{3}
\delta_{BR} = -\frac{L^2}{12 r^2}(1 - \cos \alpha) + {\cal O}(L^4)\, .
\end{equation}

The above expression is somewhat complementary to~\eqref{1} in that it is \emph{maximized} for aligned points ($\alpha = \pi$). Notice also that the embedding intuition is consistent with $\delta_{BR}$ being negative-definite because ``cutting through" some ambient space  always represents a shortcut. The negative sign in~\eqref{3} is a necessary condition for the new distance to still define a metric space. We argue, without proving, that the average metrics defined in the last section also define a metric space, in the sense that they satisfy the triangle inequality.

Notice that spatial curvature, when measured with the triangle diagnostic, becomes an ambiguous quantity in this beyond-Riemannian framework, because the correction~\eqref{3} is of the same order as the Riemannian one~\eqref{1}.
 So for instance there is a choice of the parameter $1/r^2$ such that the total discrepancy $\delta_R + \delta_{BR}$ becomes independent of the angle $\alpha$. When we measure the spatial curvature of the universe we effectively look for some $\alpha$ dependence in our geometric probes. It is intriguing to imagine that the \emph{curvature problem}, traditionally addressed by primordial inflation, could ``just not be there" in some extended geometric framework.

\section{Discussion}

In this paper we have considered averaged spacelike distances in states with relevant quantum fluctuations and found that they do not add up at large separation.   
In what follows we sketch some possible implications of these reasonings on holography and black hole evaporation. 
Then we turn to timelike and lightlike distances and discuss what it would mean, for them, to be non 
additive. The emerging ``beyond-Lorentzian" picture looks too preposterous to be omitted.

\subsection{Islands, fuzzballs and (the breakdown of) classicality}

A standard prescription to make local operators gauge invariant is that of ``dressing" them~\cite{Donnelly:2015hta,Donnelly:2016rvo}. In holographic AdS this can be achieved by defining a non-local object that contains the metric operator evaluated all the way from the boundary to the point that we want to reach. For example, a Wilson line implicitly contains the geometrical prescription of how to reach some given point of the bulk from the boundary (e.g. ``Leave the boundary at point $x^\mu$ in the orthogonal direction and proceed straight along a geodesic for 120 meters"). 

It is clear that in the presence of relevant fluctuations in the metric field the Wilson line procedure would miss its target, and define instead some non local field operator spread about an \emph{extended region of space}. This is a vague statement which our observers can make more precise, as detailed in  App.~\ref{appen}. We are saying that a Wilson line operator will hit different observers with different calculable probabilities,  rather than picking up just one of them.

We argue that something similar might be happening inside real Lorentzian spacetimes when Euclidean calculations start being dominated by replica wormholes~\cite{Almheiri:2020cfm}. The ``island" that at some point forms  inside an evaporating black hole~\cite{Almheiri:2020cfm} is disconnected from the boundary, separated by a region which, we argue, could be characterized by relevant quantum fluctuations in the metric field. As a consequence, the local operators inside the island cannot be dressed by any simple Wilson line procedure and require a much more complicated superposition of boundary operators. 

Once the island forms, reaching any given observer with a simple ``classical prescription" becomes impossible. As a consequence, the picture of the hole from the boundary looks blurry, like a ``fuzzball". However, and contrary to what the fuzzball paradigm~\cite{Mathur:2005zp} seems to suggest, the picture looks extremely smooth as we get close to the hole. As argued in Sec.~\ref{2.1}, the effects of ``highly quantum" gravitational states are locally invisible. So nothing dramatic would happen at the horizon of the black hole.

One main takeaway of the present paper is that  the breakdown of classicality has no local consequences. The consequences are relegated to the infrared, in the sense that beyond-Riemannian effects build up at large mutual separations and cannot be detected with local experiments.
 The anomalous \emph{chord}-like distance emerging from our picture---with its property of being ``locally flat" around each point---is an example of this  phenomenon. 
Our fluid of observers could safely populate the region very close to the black hole%\footnote{Such a region should be suitably defined, for instance by trying to generalize quantum mechanically the concept of closed trapped surface.}
and inside it, and such regions will look smooth and locally flat for each one of them.

%It could be that these novel points of view could help addressing e.g. the difficulties recently raised in~\cite{Geng:2021hlu}. A more in depth analysis of these issues  is left for future work. 

\subsection{Beyond-Lorentzian?}\label{sec_bl}

\begin{figure}[h]
\captionsetup{width=.9\linewidth}
\begin{center}
\includegraphics[width=11cm]{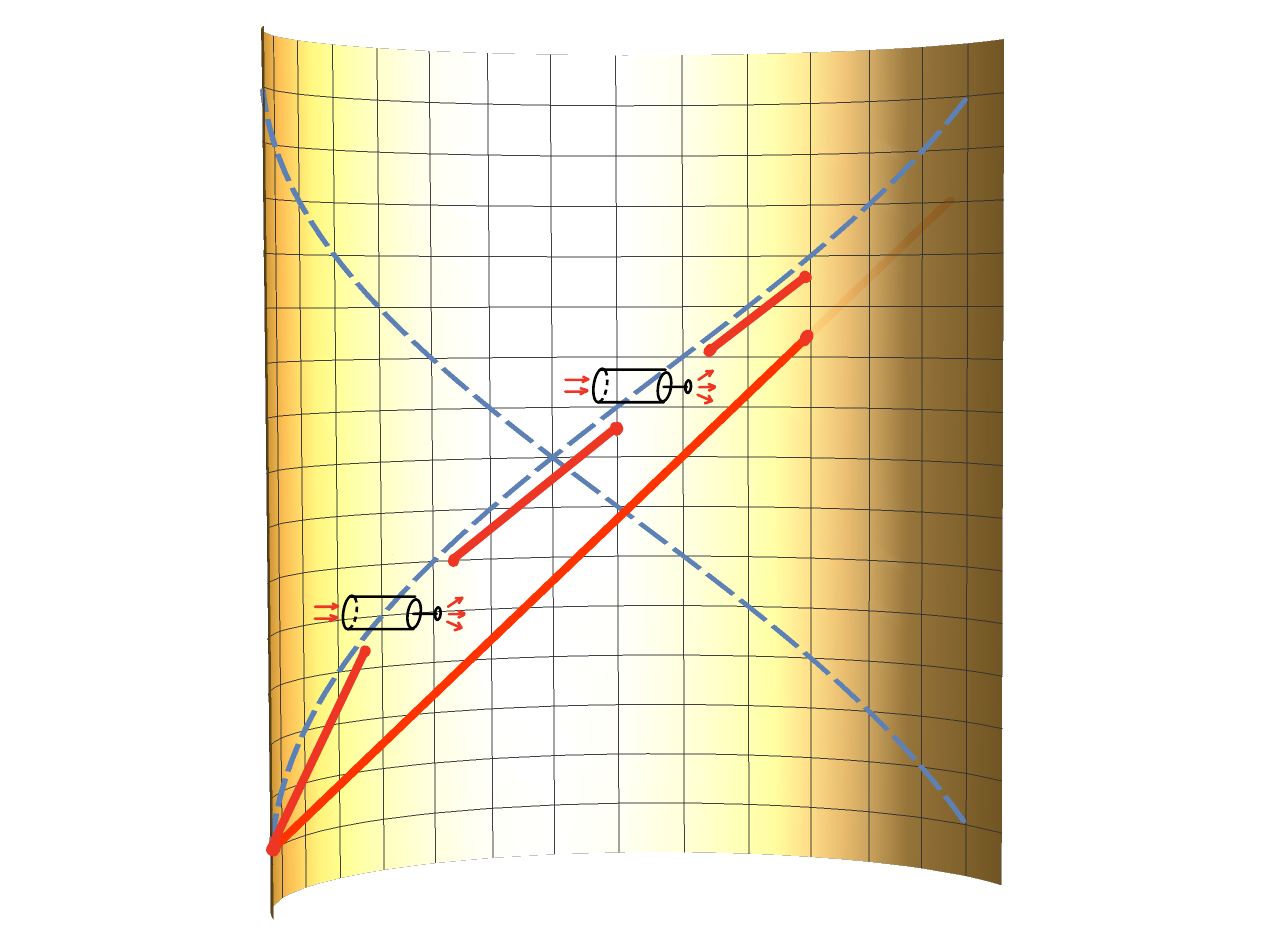}
\end{center}
\caption{Time moves along the vertical axis and space is horizontal in this portrait of an embedded Lorentzian manifold. Between emission and absorption coherent photons states connect events that are at null separation with respect to the beyond-Lorentzian distance. They ``jump" inside the ambient space. The classical lightcone structure of general relativity emerges in the limit in which photons are continuously absorbed and reemitted in the same direction.\label{figcyl2}}
\end{figure}

 Thanks to the time-field $T(x^\k)$ introduced in Sec.~\ref{sec_timefield} we have managed to deal with spacelike distances ``along some given time-slice" and keep away from the subtleties related with time evolution in quantum gravity. 
We see no reason why the non-additivity property of spacelike distances should disappear once time evolution will be properly taken into account.  
If anything, searching for spacelike geodesics not restricted to the $T = const.$ surface should make the picture even richer. Moreover, we generally expect this non-additivity to leak also in the timelike and lightlike directions. At least this is the extremely naive picture that we get if we add an orthogonal straight time direction to an embedded Riemannian manifold to make it an embedded \emph{Lorentzian} manifold (Fig.~\ref{figcyl2}).

Timelike distances represent the proper time of some observer---how could they possibly be non-additive? After all, time intervals just add up to one another as much as the minute hand of a clock continuously probes the circumference of the clock itself. In this sense, time intervals are additive by construction, at least if the clock is macroscopic and we keep watching it. But suppose instead that we keep track of proper time by (quantum-) measuring a system that evolves coherently without external interactions until the measurement takes place. For instance we could consider the relative phase of the wavefunction of a massive particle beam-splitted into two slightly different paths. 
Between the instant when the particle is beam splitted and the one when the relative phase is measured there is no such thing as ``keeping watching the clock". 

So a way to make sense of non-additive time intervals seems to be that of associating them with subsystems that maintain quantum coherence and do not entangle with anything else. And to argue that, 
on the opposite, classical spacetime (say, the one equipped with the effective metric ${\cal G}_{IJ}$ of eq.~\eqref{4.16}, which only accounts for distances \emph{locally}) is the domain of decoherence and dynamical entanglement. A macroscopic clock measures proper time as given by the metric line element integrated along the geodesic. The ``quantum-coherent" clock described above measures instead the beyond-Lorentzian time interval between the events ``preparation of the system" and ``measurement".

Even more strikingly,  the Lorentzian distances inherited from the ambient space can change the causal structure.  Null geodesics of the ambient manifold can connect events that are spacelike separated for the embedded one (Fig.~\ref{figcyl2}). Again, these ``jumps" could characterise the behavior of a photon that maintains quantum coherence between its production and its absorption.
On the opposite, the lightcone structure of general relativity is defined by photons that continuously decohere with the environment. 
 
The locally flat limit of chord distances is respected also in Lorentzian signature so 
these ``superluminal" effects would not be observable locally and would become important only at some large separation set, say, by a curvature scale (e.g. a CMB photon produced at recombination and collected by the Planck satellite. See also~\cite{Ishihara:2000nf} for related ideas).   The healthy physicist should immediately become suspicious when it comes to superluminality. On the other hand,  the leak of information from an evaporating black hole should probably come to terms with violations of causality in a way or another~\cite{Mathur:2009hf,Raju:2020smc} and a beyond-Lorentzian geometry could stage this process. 

{\bf Note added:} After this paper's first appearance on the archives I became aware of the interesting work~\cite{Omiya:2021olc}. In the context of double-holography, these authors find violations of causality not dissimilar to those just discussed, which build up in the infrared, or at large separations. The embedded-manifold picture capturing the behavior of BR-distances (e.g.~Fig.~\ref{figcyl2}) seems to have a precise analogue in the \emph{bulk picture} of the AdS boundary with an end-of-the-world brane attached to it.  
It will be interesting to explore the possible connections between the geometrical anomalies found here and the ``IR-sensitive non-locality" of~\cite{Omiya:2021olc}.

\section*{Acknowledgements} I am particularly grateful to Alberto Nicolis for endless discussions about many of the topics touched upon in this paper. I also acknowledge useful exchanges with Raphael Bousso, Sergio Cacciatori, Tom Hartman, Federico Re, Alberto Verga and Kenza Zeghari. I thank Zixia Wei for a detailed explanation of his work~\cite{Omiya:2021olc} and for making me aware of other relevant references. 

\appendix
\section{The AdS boundary and the rest} \label{appen}

In asymptotically Anti de Sitter (AdS) space the boundary can efficiently serve as an ``observer": a most reliable one, because the fluctuations of the metric dynamically vanish on it. When discussing geometric constructions in AdS that anchor to its boundary it might be useful to specify also \emph{what we anchor to the other side}. To this purpose we populate the AdS bulk with observers. Let us consider the observers ``test particles" of Sec.~\ref{sec_2new}: $A$, $B$, $C$ etc.

With minor adjustments we can extend the definition of distance operator to include the boundary, so that quantities like e.g. $\hat d(A,{\rm Boundary})$ can be evaluated on some state. As for the distance operator among pointlike observers, we define $\hat d(A,{\rm Boundary})$ by its action on a classical basis like in eq.~\eqref{classbasis}. On each element of this basis a straightforward geometrical computation produces a definite eigenvalue for the distance operator. While doing this, we have to pay attention to consistently renormalizing the (otherwise, infinite) distance, which we leave implicit in what follows. Moreover, when we minimize the length of all possible curves to obtain the geodesic, one should also vary over all possible boundary points.

In this setup several questions can be asked. What is the distance from $A$ to the boundary? This is a quantum mechanical quantity with an average (expressed by the appropriate generalization of~\eqref{avedist}) an uncertainty etc. 

Alternatively, we can consider a typical ``Wilson line" geometrical prescription to define a point in the bulk. For instance
\begin{quote}
\emph{Prescription}: Leave the boundary at point $y^\mu$ in the orthogonal direction and proceed along a geodesic for 120 meters.
\end{quote}
And ask: will such a prescription define a point in the bulk? The problem with this question is that if we are dealing with superpositions of geometries it is not even clear what it means ``a point in the bulk". Our observers, however, can help setting this question on a slightly firmer basis.

To each element of the classical basis the \emph{Prescription} assigns a coordinate point unambiguously, 
\begin{equation}
| h_{ij}(x^\k), x^i_{A},\dots \rangle \quad \longrightarrow \quad X^i[h], 
\end{equation}
which is a functional of the corresponding metric and does not depend on the observers' positions. 
The probability amplitude that $(i)$ the metric is in some configuration $h_{ij}(x^k)$, $(ii)$ the observers $B$, $C$ etc. are at positions $x_B$, $x_C$ etc. \emph{and} $(iii)$ that the \emph{Prescription} has reached observer $A$ is given by the wavefunction $\tp$ with $X^i[h]$ instead of the variable $x_A^i$. 
\begin{equation}
\tilde \Psi\left[h_{ij} (x^k), \, \ X^i[h], \ x_B^i , \ \dots \right]\, .
\end{equation}
Notice that this quantity depends on the metric ``twice". 
 We can now calculate the probability that the \emph{Prescription} has reached observer $A$ by tracing the corresponding density matrix elements over all variables except $x_A$,
%\begin{equation} \label{PA}
%P_A \ = \ \int {\cal D} \tilde h \ d^3 x_B\  d^3 x_C\dots \left|\tilde \Psi\left[\tilde h_{ij} (x^k), \, \ X^i[h], \ x_B^i , \ \dots \right]\right|^2\, .
%\end{equation} 
\begin{equation} \label{PA}
P_A \ = \ \int {\cal D} \tilde h \ d^3 x_B\  d^3 x_C\dots \bigg| \tilde \Psi\left[ h_{ij} (x^k), \, \ X^i[h], \ x_B^i , \ \dots \right]\bigg| ^2\, .
\end{equation} 
In general, several observers will be hit by the \emph{Prescription} with non vanishing probability. 
This means that, in the presence of relevant fluctuations of the metric, \emph{Prescription} will not define anything ``local", in the operational sense of ``being attached to some given observer". 

Making the same operation in different coordinate systems could also be interesting. In unitary gauge, by definition, there is no uncertainty in the positions of the observers: $\tp$ is delta-like in the variables $x_A$, $x_B$ etc. This means that all the uncertainty about  \emph{Prescription}  reaching or not some observer is contained in the metric fluctuations, and builds up at large separations from the boundary. 
On the opposite, we can go to Fefferman-Graham (FG) coordinates where the line element reads
\begin{equation}
ds^2 = \frac{dz^2 + \gamma_{\mu \nu}(z,y) \ dy^\mu dy^\nu }{z^2}.
\end{equation}
These coordinates are built in such a way that \emph{Prescription} always produces the same result, 
\begin{equation}
X^i_{FG} = \left(z, y^i\right),
\end{equation}
independently of the metric $\gamma_{ij}$. In particular, $z$ is always the exponential of the distance from the boundary in these coordinates. In FG-gauge, when we calculate the probability of reaching some observer like in~\eqref{PA}, the fluctuations of the metric become irrelevant. The uncertainty  is now directly encoded in the wave\emph{function} of the observer's position.

FG coordinates clearly bear similarities with the normal coordinates of Sec.~\ref{sec_normal}---in normal coordinates the distance from \emph{the origin} trivially becomes the Euclidean one. The boundary of AdS, as opposed to our generic observer, has the further advantage that the metric fluctuations vanish in its vicinity. So the uncertainty in observers' positions of the FP-wavefunction should also vanish for the observers close the boundary. In other words, we do not expect for those observers any uncertainty of quantum gravitational origin, i.e. other than the usual quantum mechanical indeterminacy of the position, which is negligible for macroscopic objects. 

\renewcommand{\baselinestretch}{1}\small
\bibliographystyle{ourbst}
%\bibliography{replicaBib}
\bibliography{references}
\end{document}